\documentclass[12pt,preprint]{aastex}
\newcommand{\zl}{$z$$<$$1$}
\newcommand{\za}{$z$$\sim$}
\newcommand{\zg}{$z$$>$$1$}

\newcommand{\Iab}{$I_{814}$(AB)}
\newcommand{\U}{$U_{300}$}
\newcommand{\B}{$B_{450}$}
\newcommand{\V}{$V_{606}$}
\newcommand{\I}{$I_{814}$}
\newcommand{\zphot}{$z_{phot}$}

\newcommand{\oii}{[\ion{O}{2}]$\lambda$3727}

\slugcomment{Accepted for publication in AJ}

\shorttitle{Redshifts in the HDFS}
\shortauthors{Sawicki \& Mall\'en-Ornelas}

\begin{document}


\title{Redshifts in the Hubble Deep Field South 
\altaffilmark{1,2}}

\author{Marcin Sawicki} 
\affil{
Dominion Astrophysical Observatory, 
Herzberg Institute of Astrophysics,
National Research Council, 
5071~West Saanich Road, 
Victoria, B.C., V9E 2E7, 
Canada; 
and California Institute of Technology, MS 320-47, Pasadena, CA~91125
}
\email{marcin.sawicki@nrc.gc.ca}

\author{Gabriela Mall\'en-Ornelas} 
\affil{
Harvard-Smithsonian Center for Astrophysics, 
60 Garden St., 
MS-15,
Cambridge, MA~02138; 
Princeton University Observatory, 
Peyton Hall, 
Princeton, NJ~08544; 
and Pontificia Universidad Cat\'olica de Chile, 
Casilla 306, 
Santiago 22, 
Chile}
\email{gmalleno@cfa.harvard.edu}

\altaffiltext{1}{Based on observations made with the NASA/ESA Hubble
Space Telescope, obtained at the Space Telescope Science Institute,
which is operated by the Association of Universities for Research in
Astronomy, Inc., under NASA contract NAS 5-26555.}

\altaffiltext{2}{Based on observations collected with the Very Large
Telescope at the European Southern Observatory, Chile, as part of ESO
programs 65.O-0674 and 69.A-0416}

\begin{abstract}
We present a catalog of 97 spectroscopic redshifts of \zl\ galaxies in
the Hubble Deep Field South (HDFS) and its Flanking Fields (FFs). In
the HDFS proper, we observed approximately half the galaxies brighter
than \Iab=24 and obtained redshifts for 76\% of them.  Targets in our
HDFS sample were pre-selected to be at \zl\ based on photometric
redshifts, while in the FFs a simple magnitude cut was used.  The
photometric redshift pre-selection in the HDFS resulted in a
spectroscopic success rate that is significantly higher than in the
FFs, where no pre-selection was applied.  The RMS precision of our
redshift measurements, determined from repeat observations, is $\delta
z$=0.0003.  We present the photometry and redshifts for the 97 objects
for which we secured spectroscopic redshifts and describe the basic
properties of this sample.
\end{abstract}


\section{INTRODUCTION}

Redshifts of distant galaxies are a fundamental tool in observational
cosmology.  Via Hubble's law, redshifts tell us distances, and thereby
let us calculate physical properties of distant objects, such as their
sizes and luminosities.  Moreover, redshifts allow us to compute
lookback times, thus letting us use the telescope as a time machine by
telling us the cosmic epoch at which we are observing the objects of
interest.

The Hubble Deep Field South (HDFS; Williams et al.\ 2000) contains
some of deepest and most extensive multi-wavelength imaging ever
carried out in the southern hemisphere (Casertano et al.\ 2000,
Labb\'e et al.\ 2003).  However, while over 150 spectroscopic
redshifts have been measured in the area of the {\it northern} Hubble
Deep Field (Cohen et al.\ 1996; Steidel et al.\ 1996; Lowenthal et
al.\ 1997), to date only a relatively small number of HDFS
spectroscopic redshifts has been secured (46 redshifts, including work
in preparation, of which 29 are at \zl; see Vanzella et al.\ 2002).
In the present paper, we present a catalog of spectroscopic redshifts
in the HDFS and its Flanking Fields which we obtained as part of our
study of faint galaxies at moderate redshift, $z$$\approx$0.5.  To
permit a study of the internal kinematics of these galaxies (G.\
Mall\'en-Ornelas \& M.\ Sawicki, in preparation), we obtained
moderately high dispersion spectrosopy, while to maximize our yield,
we pre-selected HDFS spectroscopic targets (though not those in the
FFs) by using photometric redshifts to exclude from the spectroscopic
sample galaxies likely to be at \zg.  Because of this pre-selection
procedure, our spectroscopic sample should not be regarded as unbiased
and we encourage the reader to consider the possible implications of
this pre-selection bias should they use our catalog in their work;
with this caveat, we believe that the redshift catalog we present will
prove useful in studies of galaxy formation and evolution, as well as
other follow-up work on the HDFS.

Our photometry, photometric redshifts, and target selection procedures
are described in \S\S~\ref{photometry} and \ref{targetselection}.  The
spectroscopic observations and reductions are described in
\S\S~\ref{spectroscopy} and \ref{redshifts}, respectively.  A basic
characterization of the resulting spectroscopic redshift sample is
given in \S~\ref{discussion}.  The relevant photometric and
spectroscopic data are compiled in Table~\ref{data.tab}.

\section{THE DATA}

\subsection{Photometry}\label{photometry}

Our photometric catalog was generated in a manner nearly identical to
that produced for the northern Hubble Deep Field by Sawicki, Lin, \&
Yee (1997); the only significant difference was that while the earlier
HDF-North work used the \V-band image for object detection, here we
used the \I-band image.  All object detection and photometry was
carried out using the PPP image analysis software (Yee 1991) on the
HDFS images which we obtained from the STScI website.
Table~\ref{data.tab} gives positions and photometry of all those
objects for which we subsequently secured spectroscopic redshifts.
Column 1 gives the object ID, while columns 2--6 give the celestial
coordinates (derived from the world coordinate system embedded in the
Version 1 images) as well as the pixel positions within the image
(HDFS or FF) in which the object is located.

\I\ ``optimal aperture magnitudes'' were computed within ``optimal
apertures'' whose sizes were optimized to give the highest S/N for
each object based on its photometric curve of growth (see Yee 1991).
These optimal aperture \I-band magnitudes were then corrected to
account for the small amount of light outside of the optimal apertures
by extrapolating the growth curves using aperture corrections based on
the growth curves of bright reference stars.  The resulting aperture
corrected magnitudes are the \I-band ``total'' magnitudes of the
objects.  The \I-band total magnitudes were placed on the AB scale
(Oke 1974) using the zero-points provided on the STScI HDFS website,
and are listed in column 7 of Table~\ref{data.tab}.  For HDFS objects,
colors were computed with reference to the \I-band via ``color
aperture magnitudes'': For each object and bandpass, the ``color
aperture'' used for computing colors was the smallest of (1) the
object's optimal aperture in that bandpass, (2) its optimal aperture
in the \I-band, and (3) an aperture of 1.55$\arcsec$ (39 pixels)
diameter.  Total magnitudes in \V, \B, and \U\ bands were then
computed by scaling the \I-band total magnitudes using the colors
computed within the color apertures; the \V, \B, and \U\ total
magnitudes are listed in columns 8--10 of Table~\ref{data.tab}.  No
\U, \B, or \V\ magnitudes are given for objects in the FFs since these
were only observed in the \I-band.

\subsection{Target selection}\label{targetselection}

In the HDFS proper, where multicolor data were available, we used
photometric redshifts to help us efficiently select targets for
spectroscopy.  We derived photometric redshifts using the same
technique as that used by Sawicki et al.\ (1997) in their study of the
evolution of the galaxy luminosity function and history of cosmic star
formation in the northern Hubble Deep Field.  This photometric
redshift technique identifies the most likely redshift of an object by
comparing its observed \U, \B, \V, \I\ fluxes with model fluxes
generated from a combination of empirical and model galaxy spectra
(for a detailed description of this procedure see Sawicki et al.\
1997).  The resultant photometric redshifts of HDFS galaxies are
listed in column 11 of Table~\ref{data.tab}.

Target selection was done so as to maximize the yield of objects with
bright emission lines for our internal kinematics study of faint
galaxies.  Specifically, we required an emission line to be visible in
our spectroscopic wavelength window, which usually meant that
$z$$\lesssim$$0.9$.  In the HDFS proper, we gave highest priority in
slit assignment to blue galaxies with \za0.5 and
\Iab$\la$24; we also gave preference to galaxies whose major axes were
roughly perpendicular to the dispersion direction, such that slits
could placed at or near the major axis of the galaxy.  Once all
possible targets meeting these criteria in a given mask were assigned
slitlets, the remaining slitlets were allocated to objects that were
redder, or fainter, but always with $z_{phot} \lesssim 0.9$ since for
higher $z$ there would be no detectable emission lines in our
wavelength range.  In many cases, additional, serendipitous objects
fell onto slitlets and were identified a posteriori during data
reduction; if these objects yielded robust redshifts, they, too, were
included in our catalog.  Overall, however, our goal was to obtain
moderate S/N emission-line spectra for our galaxy internal kinematics
survey rather than to measure a large number of redshifts of faint
objects.  We observed 89 objects in the HDFS proper, and 104 objects
in the Flanking Fields, including both targetted and serendipitous
objects.

\subsection{Spectroscopic observations}\label{spectroscopy}

Spectroscopy was obtained at the VLT using the FORS2 spectrograph on
the Kueyen (UT2) and Yepun (UT4) 8.2m telescopes.  Observations were
carried out in visitor mode over two observing runs --- in 2000 July
27--30 (hereafter Run~1) and 2002 July 14--17 (hereafter Run~2).  Each
run was four half-nights long, although during Run 1 we lost one
half-night to telescope fault.  Atmospheric transparency was very good
during both runs, with seeing ranging over 0.4--1.4$\arcsec$ during
Run 1 and 0.7--1.7$\arcsec$ during Run 2.

The field of view of the FORS2 instrument in its standard resolution
configuration is 6.8$\arcmin \times$6.8$\arcmin$ in the MOS mode
(Run~1) and 6.8$\arcmin\times$5.7$\arcmin$ in the MXU mode (Run~2),
and so is significantly larger than the 2.6$\arcmin\times$2.7$\arcmin$
L-shaped footprint of the HDFS.  Aligning FORS2 along the diagonal of
the HDFS --- with the dispersion direction approximately NE--SW ---
ensured that approximately half of the slitlets in each mask were
available for HDFS targets with the remaining half going to objects in
the Flanking Fields.  We adopted this mask orientation whenever
possible (9 out of 13 masks in Run~1, and all 4 masks in Run~2).

During Run~1 the MXU custom slit-mask mode was not yet available.  We
therefore had to use FORS2 in MOS mode, which consists of nineteen
22.5$\arcsec$-long mechanically movable slitlets which can be
reconfigured quickly during the night.  Consequently, our observing
strategy during Run~1 was to reconfigure the MOS slit masks throughout
the night, excluding from new masks any objects which we deemed to
have already accumulated sufficiently high S/N for our kinematics
program, or which had no trace of emission lines detected in 1--2
hours of exposure time.  Most masks had 9 slits on HDFS objects and
5--9 FF objects, plus 3 additional objects outside the HDFS or its
Flanking Fields.  Many objects were observed on more than one mask.
Typical exposure times were 3600s or 4500s per mask and we observed a
total of 13 masks.  Total integration time was 14 hours.

During Run 2 we used the MXU custom slit-mask mode of FORS2, which
resulted in a substantial multiplex gain, as we were able to observe
43--50 slitlets per mask, with 22--30 of those slits allocated to HDFS
objects.  For the purpose of our internal kinematics program, we
aligned the MXU slits with the major axis of each galaxy whenever the
galaxy shape could be discerned in the HST image.  Because in the MXU
mode slit masks had to be designed and manufactured more than a day
ahead of the observations, our observing strategy was simply to
include the fainter targets in multiple slit masks, while placing the
brightest objects in only one mask each.  We observed a total of four
MXU slit masks; integration times were 1800--4200s per exposure, with
3600s being very much the norm.  We made 20 exposures with a total
integration time of 18.7 hours.

During Run 1 we used the GRIS\_600R+24 grism and order separation
filter GG435+81, with the standard resolution collimator.  With
0.5$\arcsec$-wide slits the resolution was R$\sim$2460, or 120 km/s
--- sufficient to resolve the \oii\ doublet, which has a separation of
224 km/s between components.  This allowed us to unambiguously secure
spectroscopic redshifts even in those cases where {\it only} the \oii\
feature was present in our spectral range.  During Run~2, we used
grism GRIS\_1400V+18 for two masks, and GRIS\_1200R+93 with order
separation filter GG435+81 on another two masks.  We used
custom-milled slit masks with 0.6$\arcsec$ slitlets, which gave us
spectral resolution R$\sim$3500, or $\sim$85 km/s, which clearly
resolved the \oii\ doublet.

\subsection{Data reduction and determination of redshifts}\label{redshifts}

Data reduction was done using IRAF\footnote{IRAF is distributed by the
National Optical Astronomy Observatory, which is operated by the
Association of Universities for Research in Astronomy, Inc., under
cooperative agreement with the National Science Foundation.}.  In both
Runs and 1 and 2, spectral frames were bias-subtracted, and were then
excised into individual 2-D spectra for each slitlet.  Wavelength
solutions were derived using arc lamp lines for all objects and the
spectra were re-sampled into wavelength-calibrated 2-D spectra using
IRAF tasks IDENTIFY, FITCOORDS and TRANSFORM.  The angled-slit spectra
of Run~2 were rectified column-by-column using spectral arc lamp lines
as part of this wavelength calibration and resampling procedure.  Sky
lines were subtracted from the wavelength-calibrated 2-D spectra using
IRAF task BACKGROUND.  Cosmic rays were removed following the
procedure described in Ellison, Mall\'en-Ornelas, \& Sawicki (2003).
Finally, one-dimensional spectra were extracted for all objects.
Redshifts were determined by simultaneously identifying spectral
features in the one- and two-dimensional spectra.

The primary redshift indicators were hydrogen, oxygen, and neon
emission lines (see Table~\ref{data.tab}).  In most cases where we
were able to determine redshifts, it was either by identifying
multiple emission lines, or by unambiguously resolving the \oii\
doublet.  In a small number of objects only one emission line was
clearly detected but we were able to ascertain its identity, and hence
the object's redshifts, through confirmation with absorption lines
such as \ion{Ca}{2} H and K, G-band, H$\eta$, or H$\zeta$.  Finally,
in about half a dozen cases, no emission lines were detected, but we
identified the objects' redshifts from absorption lines alone.
Columns 10 and 11 of Table~\ref{data.tab} list the spectroscopic
redshifts of our objects and the spectral features with which we
identified their redshifts.  Note that the list of spectral features
is not meant to be an exhaustive list of the features that may be
present in each spectrum, but rather an indication of the clear
features used for each redshift identification.

For all the objects listed in Table~\ref{data.tab}, redshift
identifications are robust in the sense that the spectral features
used to identify redshifts are strong and unambiguous; we feel that
there is little chance of catastrophic errors in the redshifts
presented in Table~\ref{data.tab}.  Typical uncertainty in redshift
measurement, as determined by comparing the redshifts derived from
different lines in the same spectra, is 0.0001 in redshift.  A more
robust estimate of the redshift uncertainty can be gained by comparing
redshifts derived from repeated observations of the same objects; we
observed 23 objects with at least two instrumental set-ups each and
found the RMS scatter in redshift determination to be 0.0003.  We
adopt this latter figure as the representative uncertainty in the
redshifts we present.  Overall, we observed 193 objects in the HDFS
and the Flanking Fields, including both targetted and serendipitous
ones, and secured 97 redshifts (listed in Table~\ref{data.tab}) which
we consider to be robust in the sense that they are unambiguously
identified by either multiple spectral features or the resolved \oii\
doublet.

Finally, we note that a small number of objects that the PPP
photometry package identified as separate galaxies in the HST images
were blended in the images and spectroscopy we obtained from the
ground.  Since it is impossible to reliably assign the spectroscopic
redshift to one or the other of the blended objects, we list the
blended objects at a common redshift and identify them with asterisks
in column 12 of Table~\ref{data.tab}.

\section{Discussion and Summary}\label{discussion}

\begin{figure}
\plotone{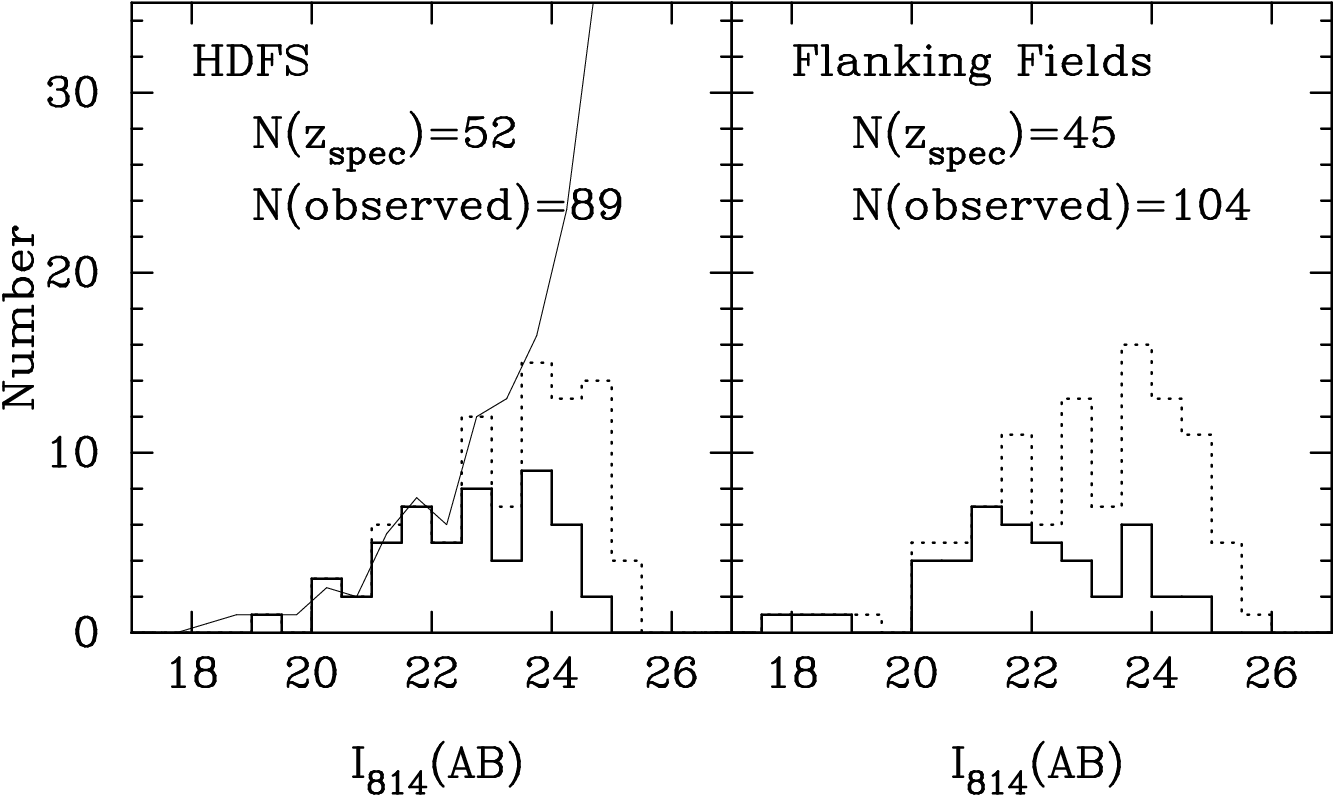}
\caption{\label{successrate_differential.fig} The success rate of
redshift determination for objects in the HDFS proper, where
photometric redshift pre-selection was used (left panel), and in the
Flanking Fields, where we did not apply any pre-selection (right
panel).  The dotted histograms show the number of objects observed in
each subsample and the solid histograms give the numbers of objects
where we secured spectroscopic redshifts.  In the left panel, the thin
solid curve shows the number counts of all HDFS galaxies {\it scaled
down by a factor of 2}: we observed approximately a half of all HDFS
galaxies to \Iab=24 and obtained redshifts for 76\% of them. }
\end{figure}

Our spectroscopic sample of \zl\ galaxies is the largest such sample
in the HDFS and its Flanking Fields.  In the HDFS proper, we observed
approximately half of all galaxies brighter than \Iab=24 (including
both targetted and serendipitous objects) and were successful in
obtaining redshifts for 76\% of them; our spectroscopic redshift
catalog accounts for about a third of all galaxies in the HDFS that
are brighter than \Iab=24 (see left-hand panel of
Fig.~\ref{successrate_differential.fig}).

\begin{figure}
\plotone{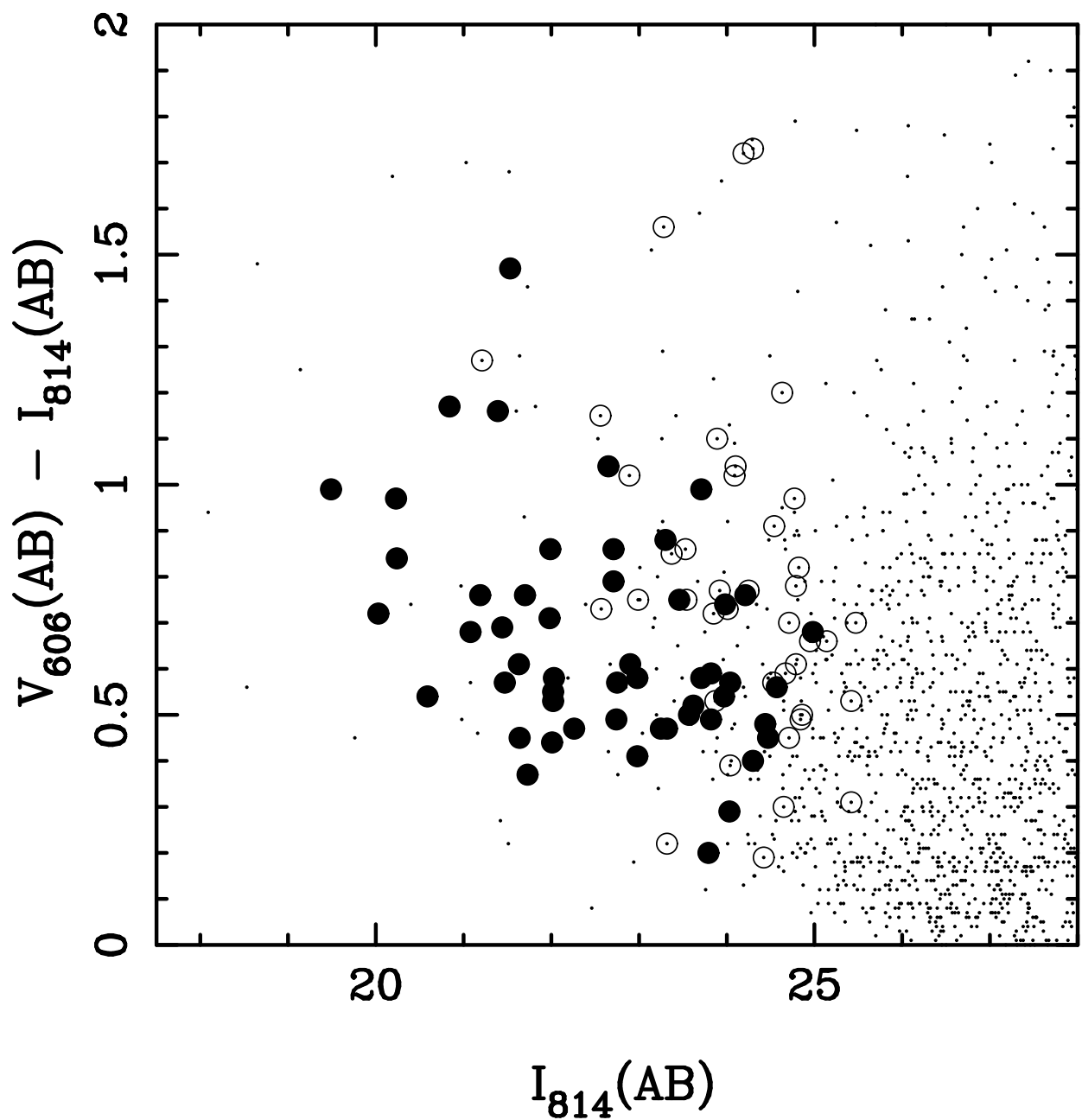}
\caption{\label{success_vs_color.fig} The colors and magnitudes of our
sample compared to the HDFS as a whole.  The small points show the
colors and magnitudes of all the HDFS objects in our photometric
catalog.  The open circles show those objects for which we obtained
spectroscopy, while the filled circles show those objects for which we
secured redshifts.  In this Figure, each blended object has been
assigned the magnitude of its brightest component object in the HST
image.}
\end{figure}

Figure~\ref{success_vs_color.fig} shows the color-magnitude diagram of
all galaxies in the HDFS proper (small points) and illustrates the
relationship of our spectroscopic sample to the underlying
color-magnitude distribution.  Large open circles identify objects
which we observed spectroscopically, including both targetted and
serendipitous ones.  Large filled circles represent galaxies for which
we were successful in securing redshifts.  Not surprisingly, we are
more successful at identifying redshifts for blue galaxies than red
ones, a reflection of the fact that blue galaxies tend to be more
strongly star-forming and consequently much more likely to have
strong, easily identifiable emission lines.

Our overall success rate for redshift identification is substantially
higher in the HDFS (52/89=58\%) than in the Flanking Fields
(44/105=43\%).  Moreover, as Figure~\ref{successrate_differential.fig}
shows, the difference in success rates for {\it faint} galaxies is
even larger: we secured redshifts for 21/49=43\% of 23$<$\Iab$<$25
galaxies in the HDFS, while in the Flanking Fields the success rate
for such faint galaxies was only 12/47=26\% --- i.e., the success rate
is 1.7 times higher in the HDFS than in the Flanking Fields.  The true
success rate for targets pre-selected with photometric redshifts is
likely to be even higher, since the HDFS target set contains a mixture
of truly pre-selected galaxies and galaxies that have been observed
because they were either allocated unused slits or serendipitously
fell into slits already assigned to other targets.  We regard the high
success rate in the HDFS, as compared with the lower success rate in
the FFs, as vindication of our approach of using photometric redshifts
to efficiently pre-select \zl\ candidates for spectroscopy.
Pre-selection is particularly important for obtaining a high yield of
{\it intrinsically faint} galaxies at \zl, since many targets towards
the faint end of our apparent magnitude limit will actually be
luminous galaxies at \zg\ with no detectable spectral features within
our wavelength range.

\begin{figure}
\plotone{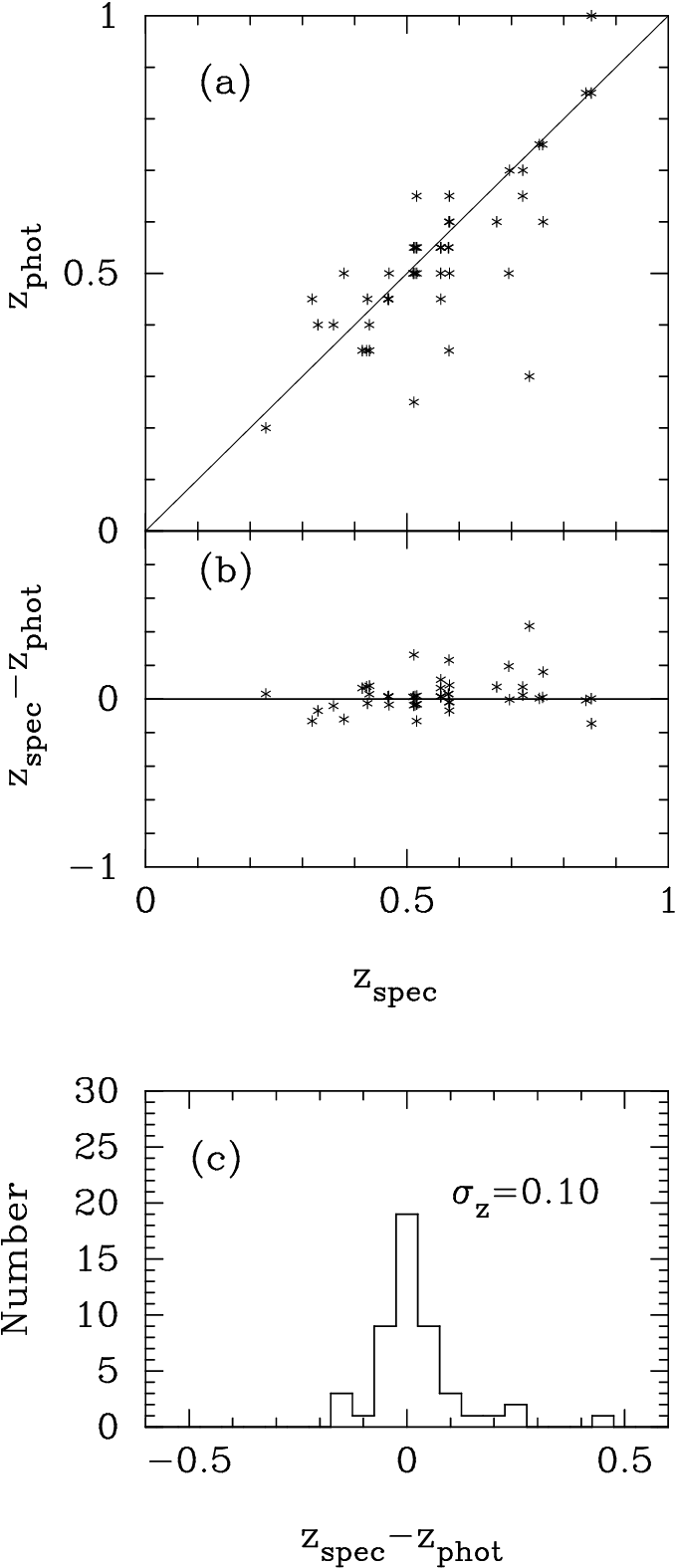}
\caption{\label{zz.fig} The photometric redshifts of those objects for
which we secured spectroscopic redshifts.  Panel (a) shows a direct
comparison between photometric and spectroscopic redshifts, and panel
(b) shows the residuals as a function of redshift.  Panel (c) shows
the histogram of the residuals.  }
\end{figure}

Figure~\ref{zz.fig} shows a comparison between spectroscopic redshifts
and our four-filter photometric redshifts in the HDFS; blended
objects, as well as the one object with a very large photometric
redshift (object \# 1551) are omitted here, leaving us with a sample
of 48 objects for this analysis.  Panels (a) and (b) show a comparison
between photometric and spectroscopic redshifts, and the redshift
residuals as a function of spectroscopic redshift, respectively.
Despite a few outliers, the agreement between photometric and
spectroscopic redshifts is very good, as is confirmed by the histogram
of the residuals, shown in panel (c).  The RMS scatter in the
residuals is $\sigma_z$=0.10, nearly identical to the scatter found by
Sawicki et al.\ (1997) using the same photometric redshift technique
in the northern HDF.  We note, however, that the spectroscopic sample
presented here does not allow for a truly blind test of photometric
redshifts as it was pre-selected {\it using} photometric redshifts.

\begin{figure}
\plotone{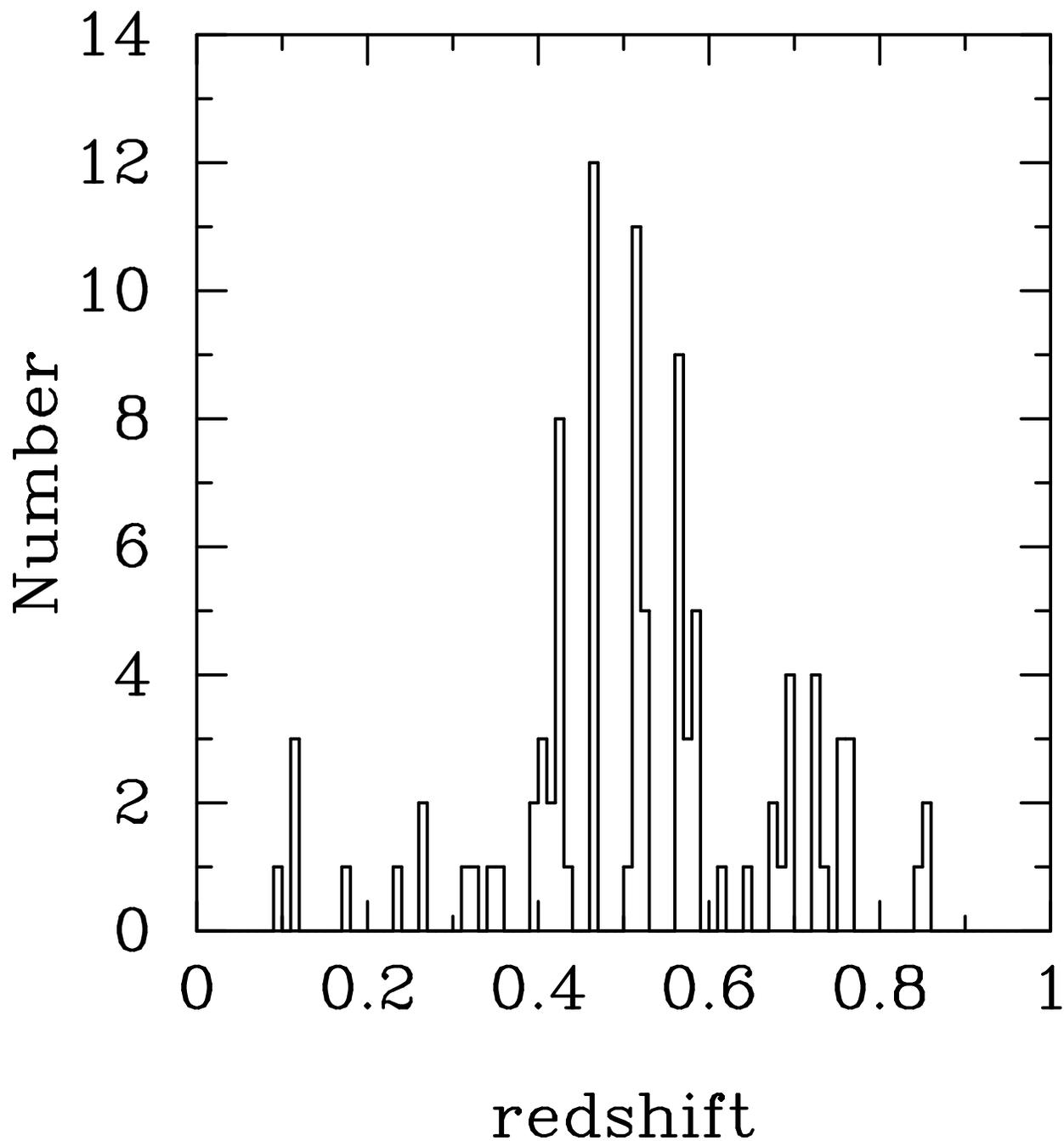}
\caption{\label{redshiftdistr_total.fig} Redshift distribution of
objects with secured spectroscopic redshifts in the HDFS and Flanking
Fields.  The bin size of the histogram is $\Delta z$=0.01.  While the
strong clustering on small scales is real, the overabundance of
objects at $z$=0.4--0.6 is due to a combination of our photometric
redshift pre-selection in the HDFS, the tuning of the FORS2 spectral
range towards detecting emission lines from galaxies in this redshift
range, and the actual redshift distribution of \Iab$\lesssim$24
galaxies.}
\end{figure}

The observed redshift distribution in the HDFS and the FFs is shown in
Figure~\ref{redshiftdistr_total.fig}.  The large number of objects at
$z$=0.4--0.6 results from the combination of pre-selection of
\zphot$\sim$0.5 galaxies in the HDFS, our tuning of the FORS2
spectral range towards detecting emission lines --- especially \oii\
--- at these redshifts, and the actual redshift distribution of
galaxies with \Iab$\lesssim$24.  The observed redshift distribution
shows strong clustering in redshift space.  This is particularly
evident in the $z$=0.4--0.6 redshift interval, which contains enough
galaxies to identify four or five distinct redshift spikes.  Such
clustering in redshift is common in pencil-beam surveys and has also
been noted in the northern HDF (Cohen et al.\ 1996).

We summarize this paper as follows:

\begin{enumerate}

\item The \zl\ spectroscopic redshift catalog presented here is the
largest redshift catalog available for the HDFS and its Flanking
Fields; it contains 52 redshifts in the HDFS and 45 in the Flanking
Fields.

\item All redshifts presented here can be considered robust, since
they have been identified using strong, unambiguous spectral features.
The typical accuracy of these redshifts is $\delta$$z$=0.0003, as
measured from repeat observations.

\item We stress that the catalog is biased towards \za0.5 galaxies,
particularly so in the HDFS proper, where we used photometric
redshifts to pre-select targets for spectroscopy.

\end{enumerate}

The reader is encouraged to use this catalog for his/her purposes, but
is cautioned to be mindful of the effects that the biases introduced
by photometric redshift pre-selection and limited spectral range of
our spectra may introduce into the analysis that is being pursued.


During part of this work, M.~S. was supported by a fellowship from the
Natural Sciences and Engineering Research Council (NSERC) of Canada
and by NSF grant AST 96-18686.  M.~S. acknowledges support for visits
to Chile from Proyecto Fondecyt de Incentivo a la Cooperaci\'on
Internacional 2000 7000529.  G.~M.-O. is currently funded by a Clay
Fellowship at the Smithsonian Astrophysical Observatory.
G.~M.-.O. acknowledges support from Fundaci\'on Andes and Proyecto
Fondecyt Regular 2000 1000529 during part of this project.  The
authors thank the staff at Paranal Observatory for outstanding support
and Bev Oke for help with some of the spectra.


\clearpage



\begin{figure}
\plotone{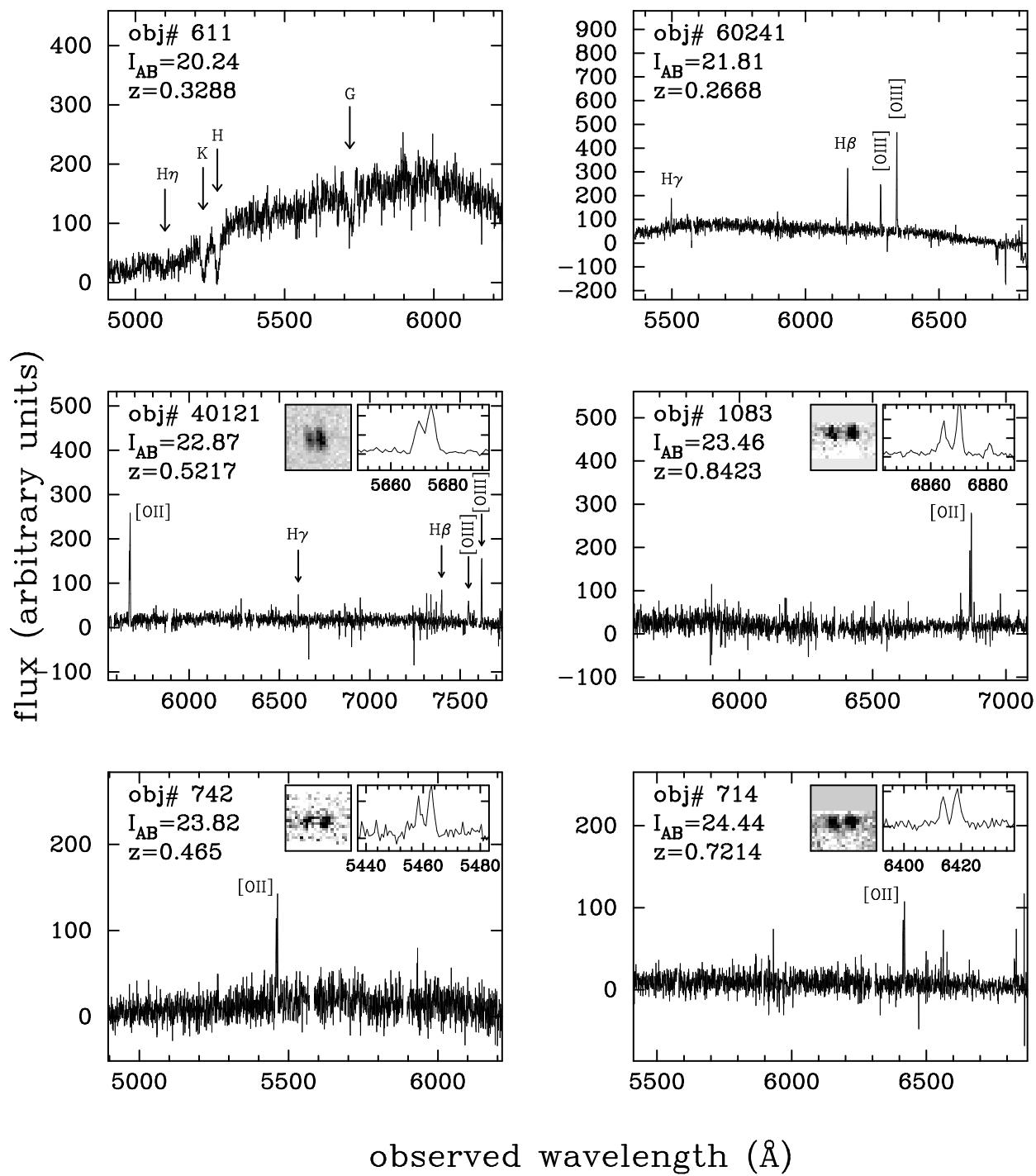}
\caption{\label{example_spectra.fig} Examples of our spectra, The
pabels are in order of increasing \I\ magnitude.  The strongest sky
subtraction residuals were masked out, although weaker ones were
retained as were single-pixel noise spikes.  Spectral features used
for redshift identifications are indicated.  Inset panels show on
expanded scale, both in 2- and 1-D, regions around the \oii\ doublet.}
\end{figure}

\twocolumn
\onecolumn



\begin{deluxetable}{llllrrcccclll}
\tabletypesize{\scriptsize}
\tablewidth{0 in}
\tablecaption{\label{data.tab} HDFS objects with spectroscopic redshifts } 
\rotate 
\tablehead{ 
\colhead{ID} & 
\colhead{RA(2000)\tablenotemark{a}} & 
\colhead{Dec(2000)\tablenotemark{a}} & 
\colhead{Field} & 
\colhead{x(pix)} & 
\colhead{y(pix)} & 
\colhead{$I_{814}(AB)$} & 
\colhead{$V_{606}(AB)$} & 
\colhead{$B_{450}(AB)$} & 
\colhead{$U_{300}(AB)$} & 
\colhead{$z_{phot}$} & 
\colhead{$z_{spec}$\tablenotemark{b}} & 
\colhead{spectral features used\tablenotemark{c,d}}
}
\startdata
30438 &  22:33:18.698 &  $-$60:34:37.16 &  FF3 & 1415 & 1667  &  18.51$\pm$0.01 &  ...           &  ...           &  ...            & ...  &  0.0990&  He5876 N6548 H$\alpha$ N6583 \\ 
 1551 &  22:32:53.838 &  $-$60:32:13.13 & HDFS & 2446 & 3592  &  23.79$\pm$0.01 & 23.99$\pm$0.01 & 24.22$\pm$0.01 & 25.06$\pm$0.03  & 1.85 &  0.1148&  O4959 O5007 H$\beta$ \\ 
60429 &  22:32:56.822 &  $-$60:36:47.80 &  FF6 & 1537 & 1634  &  17.73$\pm$0.01 &  ...           &  ...           &  ...            & ...  &  0.1182&  O4959 O5007 H$\beta$ H$\gamma$ \\ 
40329 &  22:33:20.584 &  $-$60:33:39.52 &  FF4 & 2637 & 1439  &  20.70$\pm$0.01 &  ...           &  ...           &  ...            & ...  &  0.1186&  O4959 O5007 H$\beta$ \\ 
30668 &  22:33:06.757 &  $-$60:33:43.27 &  FF3 & 3185 & 2751  &  20.13$\pm$0.01 &  ...           &  ...           &  ...            & ...  &  0.1726&  O4959 O5007 H$\beta$ H$\gamma$ \\ 
  583 &  22:33:05.878 &  $-$60:33:43.33 & HDFS &  199 & 1348  &  21.73$\pm$0.01 & 22.10$\pm$0.01 & 22.67$\pm$0.01 & 23.56$\pm$0.04  & 0.20 &  0.2301&  O4959 O5007 H$\beta$ H$\gamma$ \\ 
60241 &  22:33:01.580 &  $-$60:37:13.79 &  FF6 &  833 & 1111  &  21.81$\pm$0.01 &  ...           &  ...           &  ...            & ...  &  0.2668&  O4959 O5007 H$\beta$ H$\gamma$ \\ 
90549 &  22:32:54.810 &  $-$60:37:47.75 &  FF9 & 1687 & 2379  &  18.41$\pm$0.01 &  ...           &  ...           &  ...            & ...  &  0.2668&  O4959 O5007 \\ 
 1664 &  22:32:50.802 &  $-$60:31:59.70 & HDFS & 3011 & 3924  &  24.03$\pm$0.01 & 24.32$\pm$0.01 & 24.92$\pm$0.02 & 25.27$\pm$0.04  & 0.45 &  0.3187&  O3727 Ne3869 H$\gamma$ \\ 
  611 &  22:33:05.772 &  $-$60:33:41.45 & HDFS &  219 & 1395  &  20.24$\pm$0.01 & 21.08$\pm$0.01 & 22.60$\pm$0.02 & 24.31$\pm$0.04  & 0.40 &  0.3288&  H K G H$\eta$ \\ 
30048 &  22:33:17.750 &  $-$60:35:34.51 &  FF3 & 1556 &  514  &  21.88$\pm$0.02 &  ...           &  ...           &  ...            & ...  &  0.3442&  O4959 O5007 H$\beta$ \\ 
 1029 &  22:32:48.998 &  $-$60:33:09.41 & HDFS & 3329 & 2172  &  24.30$\pm$0.01 & 24.70$\pm$0.01 & 25.58$\pm$0.03 & 26.04$\pm$0.08  & 0.40 &  0.3595&  O3727 \\ 
80291 &  22:33:32.687 &  $-$60:34:44.97 &  FF8 & 3549 & 1229  &  20.03$\pm$0.01 &  ...           &  ...           &  ...            & ...  &  0.3938&  H K G H$\eta$ \\ 
50123 &  22:33:12.676 &  $-$60:36:28.60 &  FF5 &  748 &  732  &  22.61$\pm$0.03 &  ...           &  ...           &  ...            & ...  &  0.3947&  O3727 O4959 O5007 H$\beta$ \\ 
50649 &  22:32:59.700 &  $-$60:34:47.08 &  FF5 & 2670 & 2775  &  23.71$\pm$0.05 &  ...           &  ...           &  ...            & ...  &  0.4049&  O3727 \\ 
50681 &  22:32:59.625 &  $-$60:34:36.88 &  FF5 & 2681 & 2980  &  22.20$\pm$0.02 &  ...           &  ...           &  ...            & ...  &  0.4052&  O3727 H$\beta$ H K \\ 
20056 &  22:32:41.604 &  $-$60:35:15.13 &  FF2 & 2676 &  502  &  20.15$\pm$0.01 &  ...           &  ...           &  ...            & ...  &  0.4055&  O3727 O5007 H$\beta$ H$\gamma$ \\ 
50100 &  22:33:02.409 &  $-$60:36:32.06 &  FF5 & 2269 &  664  &  23.68$\pm$0.04 &  ...           &  ...           &  ...            & ...  &  0.4119&  Ne3869 H$\beta$ H$\delta$ H$\gamma$ \\ 
 1381 &  22:33:00.237 &  $-$60:32:34.03 & HDFS & 1257 & 3078  &  20.59$\pm$0.01 & 21.13$\pm$0.01 & 22.22$\pm$0.02 & 23.30$\pm$0.11  & 0.35 &  0.4147&  O3727 O5007 H$\beta$ H$\gamma$ \\ 
20085 &  22:32:44.006 &  $-$60:35:10.60 &  FF2 & 2320 &  593  &  21.69$\pm$0.02 &  ...           &  ...           &  ...            & ...  &  0.4220&  O3727 \\ 
20149 &  22:32:39.748 &  $-$60:35:01.80 &  FF2 & 2951 &  770  &  20.61$\pm$0.01 &  ...           &  ...           &  ...            & ...  &  0.4222&  H K G \\ 
30257 &  22:33:16.803 &  $-$60:35:01.54 &  FF3 & 1696 & 1177  &  24.59$\pm$0.07 &  ...           &  ...           &  ...            & ...  &  0.4228&  O4959 O5007 H$\beta$ \\ 
  743 &  22:32:58.242 &  $-$60:33:31.50 & HDFS & 1614 & 1633  &  22.01$\pm$0.01 & 22.45$\pm$0.01 & 23.33$\pm$0.03 & 24.11$\pm$0.04  & 0.35 &  0.4229&  O3727 \\ 
20277 &  22:32:43.392 &  $-$60:34:41.85 &  FF2 & 2411 & 1171  &  21.73$\pm$0.01 &  ...           &  ...           &  ...            & ...  &  0.4233*&  O3727 H K G H$\eta$ \\ 
20279 &  22:32:43.493 &  $-$60:34:41.70 &  FF2 & 2396 & 1174  &  20.09$\pm$0.01 &  ...           &  ...           &  ...            & ...  &  0.4233*&  O3727 H K G H$\eta$ \\ 
20274 &  22:32:43.466 &  $-$60:34:42.10 &  FF2 & 2400 & 1166  &  21.67$\pm$0.01 &  ...           &  ...           &  ...            & ...  &  0.4233*&  O3727 H K G H$\eta$ \\ 
  100 &  22:32:50.955 &  $-$60:34:15.39 & HDFS & 2952 &  520  &  23.97$\pm$0.01 & 24.51$\pm$0.01 & 25.45$\pm$0.03 & 26.14$\pm$0.08  & 0.45 &  0.4245&  O3727 \\ 
  501 &  22:32:57.881 &  $-$60:33:49.13 & HDFS & 1677 & 1190  &  23.25$\pm$0.01 & 23.72$\pm$0.01 & 24.84$\pm$0.03 & 25.31$\pm$0.08  & 0.40 &  0.4280&  O3727 O5007 \\ 
  962 &  22:33:01.886 &  $-$60:33:16.46 & HDFS &  943 & 2016  &  21.63$\pm$0.01 & 22.24$\pm$0.01 & 23.23$\pm$0.02 & 24.37$\pm$0.04  & 0.35 &  0.4284&  O3727 \\ 
50190 &  22:33:11.325 &  $-$60:36:17.43 &  FF5 &  948 &  957  &  23.63$\pm$0.05 &  ...           &  ...           &  ...            & ...  &  0.4334&  O3727 \\ 
20462 &  22:32:45.678 &  $-$60:34:18.17 &  FF2 & 2072 & 1647  &  21.46$\pm$0.01 &  ...           &  ...           &  ...            & ...  &  0.4608&  O3727 H$\gamma$ \\ 
20476 &  22:32:46.076 &  $-$60:34:16.48 &  FF2 & 2013 & 1681  &  21.28$\pm$0.01 &  ...           &  ...           &  ...            & ...  &  0.4609&  O3727 H$\gamma$ \\ 
   93 &  22:32:45.665 &  $-$60:34:15.47 & HDFS & 3930 &  509  &  24.67$\pm$0.04 & 25.26$\pm$0.03 & 26.74$\pm$0.12 & 27.04$\pm$0.34  & 0.45 &  0.4621*&  O3727 \\ 
  102 &  22:32:45.634 &  $-$60:34:14.83 & HDFS & 3936 &  525  &  24.57$\pm$0.03 & 25.13$\pm$0.03 & 25.90$\pm$0.06 & 26.88$\pm$0.33  & 0.15 &  0.4621*&  O3727 \\ 
  894 &  22:33:02.761 &  $-$60:33:22.09 & HDFS &  780 & 1876  &  20.23$\pm$0.01 & 21.20$\pm$0.01 & 22.64$\pm$0.02 & 24.69$\pm$0.05  & 0.45 &  0.4642&  O3727 H$\beta$ \\ 
 1596 &  22:32:52.705 &  $-$60:32:07.16 & HDFS & 2657 & 3740  &  21.64$\pm$0.01 & 22.09$\pm$0.01 & 22.84$\pm$0.02 & 23.30$\pm$0.03  & 0.45 &  0.4643&  O3727 \\ 
  631 &  22:32:59.416 &  $-$60:33:39.54 & HDFS & 1395 & 1433  &  22.02$\pm$0.01 & 22.55$\pm$0.01 & 23.34$\pm$0.02 & 24.08$\pm$0.05  & 0.45 &  0.4644&  O3727 \\ 
50481 &  22:32:58.080 &  $-$60:35:24.68 &  FF5 & 2910 & 2019  &  21.42$\pm$0.02 &  ...           &  ...           &  ...            & ...  &  0.4645&  O3727 H$\beta$ \\ 
50493 &  22:33:00.179 &  $-$60:35:20.55 &  FF5 & 2599 & 2102  &  24.23$\pm$0.08 &  ...           &  ...           &  ...            & ...  &  0.4648&  O3727 O5007 H$\beta$ H$\gamma$ \\ 
50306 &  22:32:58.870 &  $-$60:35:58.35 &  FF5 & 2793 & 1342  &  23.84$\pm$0.05 &  ...           &  ...           &  ...            & ...  &  0.4648&  O3727 \\ 
50292 &  22:33:10.192 &  $-$60:35:59.60 &  FF5 & 1116 & 1316  &  20.98$\pm$0.01 &  ...           &  ...           &  ...            & ...  &  0.4650&  O3727 \\ 
  742 &  22:33:04.269 &  $-$60:33:31.87 & HDFS &  499 & 1633  &  23.82$\pm$0.01 & 24.41$\pm$0.01 & 25.41$\pm$0.03 & 26.37$\pm$0.12  & 0.45 &  0.4650&  O3727 \\ 
  224 &  22:32:55.247 &  $-$60:34:07.58 & HDFS & 2160 &  723  &  22.98$\pm$0.01 & 23.39$\pm$0.01 & 24.05$\pm$0.01 & 24.47$\pm$0.05  & 0.50 &  0.4656&  O3727 Ne3869 H$\delta$ H$\gamma$ \\ 
10480 &  22:33:13.915 &  $-$60:32:03.35 &  FF1 &  932 & 1703  &  22.02$\pm$0.02 &  ...           &  ...           &  ...            & ...  &  0.5006&  O3727 H$\gamma$ \\ 
 1791 &  22:32:51.962 &  $-$60:31:42.76 & HDFS & 2800 & 4351  &  24.04$\pm$0.02 & 24.61$\pm$0.01 & 25.45$\pm$0.04 & 26.16$\pm$0.24  & 0.50 &  0.5122&  O3727 \\ 
 1808 &  22:32:52.067 &  $-$60:31:40.97 & HDFS & 2781 & 4396  &  22.75$\pm$0.01 & 23.32$\pm$0.01 & 24.23$\pm$0.02 & 24.49$\pm$0.04  & 0.50 &  0.5124&  O3727 \\ 
 1793 &  22:32:54.090 &  $-$60:31:42.77 & HDFS & 2406 & 4354  &  22.71$\pm$0.01 & 23.50$\pm$0.01 & 24.47$\pm$0.02 & 25.36$\pm$0.05  & 0.55 &  0.5131&  O3727 \\ 
 1750 &  22:32:56.049 &  $-$60:31:49.04 & HDFS & 2042 & 4200  &  21.44$\pm$0.42 & 22.13$\pm$0.01 & 22.98$\pm$0.01 & 24.04$\pm$0.04  & 0.25 &  0.5134&  O3727 O5007 H$\beta$ \\ 
50055 &  22:33:09.230 &  $-$60:36:39.95 &  FF5 & 1259 &  505  &  23.93$\pm$0.02 &  ...           &  ...           &  ...            & ...  &  0.5135&  O3727 \\ 
 1802 &  22:32:50.818 &  $-$60:31:41.56 & HDFS & 3012 & 4379  &  23.71$\pm$0.01 & 24.29$\pm$0.01 & 25.06$\pm$0.03 & 25.38$\pm$0.07  & 0.55 &  0.5156&  O3727 \\ 
 1206 &  22:32:54.045 &  $-$60:32:51.69 & HDFS & 2399 & 2625  &  22.02$\pm$0.02 & 22.57$\pm$0.01 & 23.38$\pm$0.02 & 23.85$\pm$0.03  & 0.50 &  0.5158&  O3727 H$\gamma$ \\ 
 1471 &  22:32:56.156 &  $-$60:32:21.53 & HDFS & 2015 & 3385  &  24.47$\pm$0.02 & 24.92$\pm$0.01 & 25.23$\pm$0.02 & 25.48$\pm$0.05  & 0.65 &  0.5186&  O3727 Ne3869 H$\delta$ \\ 
 1387 &  22:32:57.764 &  $-$60:32:33.07 & HDFS & 1715 & 3098  &  21.98$\pm$0.01 & 22.69$\pm$0.01 & 23.59$\pm$0.01 & 24.17$\pm$0.04  & 0.55 &  0.5187&  O3727 H$\beta$ H$\delta$ H$\gamma$ \\ 
 1511 &  22:32:55.534 &  $-$60:32:17.51 & HDFS & 2131 & 3485  &  23.32$\pm$0.01 & 23.79$\pm$0.01 & 24.49$\pm$0.01 & 24.83$\pm$0.03  & 0.50 &  0.5190&  O3727 H$\beta$ \\ 
 1482 &  22:32:56.044 &  $-$60:32:20.29 & HDFS & 2036 & 3416  &  22.74$\pm$0.01 & 23.23$\pm$0.01 & 23.84$\pm$0.01 & 24.34$\pm$0.03  & 0.55 &  0.5191&  O3727 \\ 
30285 &  22:33:11.438 &  $-$60:34:58.72 &  FF3 & 2491 & 1234  &  22.43$\pm$0.02 &  ...           &  ...           &  ...            & ...  &  0.5215&  O3727 \\ 
40121 &  22:33:26.540 &  $-$60:34:12.63 &  FF4 & 1754 &  773  &  22.87$\pm$0.03 &  ...           &  ...           &  ...            & ...  &  0.5217&  O3727 O4959 O5007 H$\beta$ H$\gamma$ \\ 
90730 &  22:32:45.782 &  $-$60:36:55.35 &  FF9 & 3023 & 3433  &  22.04$\pm$0.02 &  ...           &  ...           &  ...            & ...  &  0.5271&  H K H$\eta$ H$\zeta$ \\ 
50221 &  22:33:07.722 &  $-$60:36:12.10 &  FF5 & 1482 & 1065  &  24.61$\pm$0.07 &  ...           &  ...           &  ...            & ...  &  0.5275&  O3727 \\ 
50139 &  22:33:11.765 &  $-$60:36:26.07 &  FF5 &  883 &  783  &  22.85$\pm$0.04 &  ...           &  ...           &  ...            & ...  &  0.5282&  O3727 Ne3869 H$\delta$ \\ 
  111 &  22:32:56.083 &  $-$60:34:14.21 & HDFS & 2004 &  558  &  22.03$\pm$0.02 & 22.61$\pm$0.01 & 23.25$\pm$0.03 & 23.76$\pm$0.04  & 0.55 &  0.5645&  O3727 H$\gamma$ \\ 
  646 &  22:32:53.742 &  $-$60:33:37.59 & HDFS & 2445 & 1473  &  21.47$\pm$0.01 & 22.04$\pm$0.01 & 22.77$\pm$0.02 & 23.36$\pm$0.03  & 0.50 &  0.5645&  O3727 H$\gamma$ \\ 
  366 &  22:32:52.142 &  $-$60:33:59.45 & HDFS & 2736 &  922  &  21.70$\pm$0.01 & 22.46$\pm$0.01 & 23.44$\pm$0.03 & 24.37$\pm$0.07  & 0.55 &  0.5646&  O3727 \\ 
30446 &  22:33:05.886 &  $-$60:34:35.84 &  FF3 & 3314 & 1694  &  20.71$\pm$0.01 &  ...           &  ...           &  ...            & ...  &  0.5647&  O3727 H K G \\ 
 1017 &  22:32:49.504 &  $-$60:33:11.12 & HDFS & 3235 & 2130  &  22.98$\pm$0.01 & 23.56$\pm$0.01 & 24.35$\pm$0.01 & 24.88$\pm$0.04  & 0.55 &  0.5648&  O3727 \\ 
  698 &  22:32:55.736 &  $-$60:33:33.97 & HDFS & 2077 & 1567  &  23.57$\pm$0.01 & 24.07$\pm$0.01 & 25.01$\pm$0.05 & 25.66$\pm$0.16  & 0.45 &  0.5649&  O3727 \\ 
  901 &  22:32:54.335 &  $-$60:33:20.73 & HDFS & 2339 & 1897  &  23.82$\pm$0.01 & 24.31$\pm$0.01 & 24.91$\pm$0.02 & 25.28$\pm$0.06  & 0.55 &  0.5653*&  O3727 \\ 
  909 &  22:32:54.266 &  $-$60:33:20.04 & HDFS & 2352 & 1914  &  24.04$\pm$0.01 & 24.43$\pm$0.01 & 25.26$\pm$0.02 & 26.35$\pm$0.17  & 0.25 &  0.5653*&   O3727 \\ 
10345 &  22:33:06.983 &  $-$60:32:21.34 &  FF1 & 1961 & 1342  &  21.06$\pm$0.01 &  ...           &  ...           &  ...            & ...  &  0.5656&  O3727 H K G H$\eta$ \\ 
10353 &  22:33:07.313 &  $-$60:32:20.59 &  FF1 & 1912 & 1357  &  21.21$\pm$0.01 &  ...           &  ...           &  ...            & ...  &  0.5656&  O3727 H K \\ 
  195 &  22:32:47.572 &  $-$60:34:08.59 & HDFS & 3579 &  685  &  21.19$\pm$0.01 & 21.95$\pm$0.01 & 22.86$\pm$0.02 & 23.87$\pm$0.05  & 0.55 &  0.5790&  O3727 H K H$\eta$ H$\zeta$ \\ 
 1283 &  22:32:50.905 &  $-$60:32:43.03 & HDFS & 2982 & 2837  &  20.84$\pm$0.01 & 22.01$\pm$0.01 & 23.61$\pm$0.03 & 25.70$\pm$0.13  & 0.55 &  0.5797&  H K H$\eta$ \\ 
70413 &  22:33:26.451 &  $-$60:32:23.69 &  FF7 & 3314 & 1697  &  21.68$\pm$0.01 &  ...           &  ...           &  ...            & ...  &  0.5800&  H K G \\ 
  308 &  22:32:52.235 &  $-$60:34:02.76 & HDFS & 2718 &  839  &  21.99$\pm$0.01 & 22.85$\pm$0.01 & 23.82$\pm$0.03 & 24.66$\pm$0.05  & 0.60 &  0.5804&  O3727 \\ 
  672 &  22:32:47.651 &  $-$60:33:35.87 & HDFS & 3572 & 1506  &  19.49$\pm$0.01 & 20.48$\pm$0.01 & 21.64$\pm$0.01 & 23.36$\pm$0.03  & 0.35 &  0.5807&  O3727 \\ 
 1081 &  22:32:54.024 &  $-$60:33:05.52 & HDFS & 2400 & 2278  &  22.65$\pm$0.01 & 23.69$\pm$0.01 & 25.06$\pm$0.02 & 26.17$\pm$0.08  & 0.65 &  0.5811&  O3727 H K \\ 
 1040 &  22:32:52.291 &  $-$60:33:08.31 & HDFS & 2720 & 2205  &  21.08$\pm$0.01 & 21.76$\pm$0.01 & 22.58$\pm$0.02 & 23.48$\pm$0.03  & 0.50 &  0.5815&  O3727 \\ 
  938 &  22:32:55.863 &  $-$60:33:17.71 & HDFS & 2057 & 1975  &  22.90$\pm$0.01 & 23.51$\pm$0.01 & 24.21$\pm$0.02 & 24.71$\pm$0.04  & 0.60 &  0.5817&  O3727 \\ 
50321 &  22:33:01.314 &  $-$60:35:56.45 &  FF5 & 2431 & 1380  &  24.21$\pm$0.05 &  ...           &  ...           &  ...            & ...  &  0.6131&  O3727 \\ 
  564 &  22:32:49.436 &  $-$60:33:43.80 & HDFS & 3240 & 1310  &  24.21$\pm$0.01 & 24.97$\pm$0.01 & 25.97$\pm$0.05 & 26.35$\pm$0.08  & 0.60 &  0.6475*&  O3727 \\ 
  572 &  22:32:49.318 &  $-$60:33:43.11 & HDFS & 3262 & 1327  &  24.71$\pm$0.02 & 25.41$\pm$0.02 & 26.09$\pm$0.03 & 26.23$\pm$0.07  & 0.65 &  0.6475*&  O3727 \\ 
 1561 &  22:32:55.714 &  $-$60:32:11.50 & HDFS & 2099 & 3636  &  21.53$\pm$0.01 & 23.00$\pm$0.01 & 24.95$\pm$0.02 & 27.83$\pm$0.37  & 0.60 &  0.6721&  H K \\ 
50492 &  22:32:59.963 &  $-$60:35:21.54 &  FF5 & 2631 & 2082  &  23.95$\pm$0.09 &  ...           &  ...           &  ...            & ...  &  0.6744&  O3727 \\ 
20581 &  22:32:36.221 &  $-$60:33:59.87 &  FF2 & 3474 & 2015  &  22.63$\pm$0.02 &  ...           &  ...           &  ...            & ...  &  0.6898&  O3727 \\ 
  533 &  22:33:02.490 &  $-$60:33:46.51 & HDFS &  825 & 1263  &  22.26$\pm$0.02 & 22.73$\pm$0.01 & 23.41$\pm$0.02 & 23.80$\pm$0.03  & 0.50 &  0.6949&  O3727 H$\delta$ \\ 
50467 &  22:33:00.132 &  $-$60:35:28.85 &  FF5 & 2606 & 1935  &  21.40$\pm$0.02 &  ...           &  ...           &  ...            & ...  &  0.6953&  O3727 \\ 
50482 &  22:33:02.440 &  $-$60:35:24.42 &  FF5 & 2264 & 2024  &  21.22$\pm$0.01 &  ...           &  ...           &  ...            & ...  &  0.6958&  O3727 \\ 
 1748 &  22:32:54.439 &  $-$60:31:49.09 & HDFS & 2340 & 4196  &  24.98$\pm$0.02 & 25.66$\pm$0.02 & 26.12$\pm$0.04 & 26.85$\pm$0.10  & 0.70 &  0.6963&  O3727 \\ 
30075 &  22:33:12.418 &  $-$60:35:31.00 &  FF3 & 2346 &  585  &  23.49$\pm$0.05 &  ...           &  ...           &  ...            & ...  &  0.7207&  O3727 \\ 
50506 &  22:32:59.929 &  $-$60:35:17.66 &  FF5 & 2636 & 2160  &  21.52$\pm$0.02 &  ...           &  ...           &  ...            & ...  &  0.7208&  O3727 \\ 
  714 &  22:32:48.396 &  $-$60:33:32.65 & HDFS & 3435 & 1588  &  24.44$\pm$0.01 & 24.92$\pm$0.01 & 25.31$\pm$0.02 & 25.58$\pm$0.04  & 0.65 &  0.7214&  O3727 \\ 
  412 &  22:32:48.249 &  $-$60:33:55.04 & HDFS & 3457 & 1026  &  23.30$\pm$0.01 & 24.18$\pm$0.01 & 24.95$\pm$0.02 & 26.58$\pm$0.17  & 0.70 &  0.7218&  O3727 \\ 
  607 &  22:33:03.572 &  $-$60:33:41.67 & HDFS &  626 & 1386  &  20.03$\pm$0.01 & 20.75$\pm$0.01 & 21.85$\pm$0.01 & 23.31$\pm$0.03  & 0.30 &  0.7342&  O3727 \\ 
  215 &  22:32:46.011 &  $-$60:34:07.24 & HDFS & 3868 &  716  &  23.62$\pm$0.01 & 24.14$\pm$0.01 & 24.47$\pm$0.01 & 24.76$\pm$0.03  & 0.75 &  0.7533&  O3727 Ne3869 \\ 
60715 &  22:32:48.393 &  $-$60:35:31.93 &  FF6 & 2785 & 3160  &  22.23$\pm$0.02 &  ...           &  ...           &  ...            & ...  &  0.7542&  O3727 \\ 
 1550 &  22:32:55.755 &  $-$60:32:13.54 & HDFS & 2091 & 3585  &  23.98$\pm$0.01 & 24.72$\pm$0.01 & 25.24$\pm$0.03 & 26.00$\pm$0.15  & 0.75 &  0.7594&  O3727 \\ 
50032 &  22:33:09.595 &  $-$60:36:42.78 &  FF5 & 1205 &  448  &  23.35$\pm$0.05 &  ...           &  ...           &  ...            & ...  &  0.7601*&  O3727 \\ 
50029 &  22:33:09.703 &  $-$60:36:43.03 &  FF5 & 1189 &  443  &  24.04$\pm$0.06 &  ...           &  ...           &  ...            & ...  &  0.7601*&  O3727 \\ 
 1379 &  22:32:57.989 &  $-$60:32:34.32 & HDFS & 1673 & 3067  &  21.39$\pm$0.01 & 22.55$\pm$0.01 & 24.09$\pm$0.03 & 26.03$\pm$0.62  & 0.60 &  0.7606&  O3727 \\ 
30412 &  22:33:14.333 &  $-$60:34:41.05 &  FF3 & 2062 & 1589  &  21.85$\pm$0.02 &  ...           &  ...           &  ...            & ...  &  0.7643&  O3727 \\ 
 1083 &  22:32:57.245 &  $-$60:33:05.61 & HDFS & 1804 & 2281  &  23.46$\pm$0.01 & 24.21$\pm$0.01 & 24.59$\pm$0.01 & 25.04$\pm$0.03  & 0.85 &  0.8423&  O3727 \\ 
 1218 &  22:32:45.784 &  $-$60:32:50.50 & HDFS & 3928 & 2641  &  22.71$\pm$0.01 & 23.57$\pm$0.01 & 24.05$\pm$0.01 & 25.51$\pm$0.19  & 0.85 &  0.8525&  O3727 \\ 
  180 &  22:33:01.528 &  $-$60:34:10.57 & HDFS &  997 &  657  &  23.71$\pm$0.01 & 24.70$\pm$0.01 & 25.15$\pm$0.02 & 25.75$\pm$0.06  & 1.00 &  0.8532&  O3727 \\ 
\enddata
\tablenotetext{a}{From the Version 1 images of the HDFS and the Flanking Fields}
\tablenotetext{b}{Groups of redshifts marked with asterisks indicated blended objects.}
\tablenotetext{c}{Note that the above
is not meant to be an exhaustive list of features seen in each spectrum, and simply
includes clearly-detected features used for redshift measurement.}
\tablenotetext{d}{
He5876 refers to He~I emission; 
N6548 and N6583 refer to [N~II] emission; 
O3727 refers to [O~II] emission; 
O4959 and O5007 refer to [O~III] emission; 
Ne3869 refers to [Ne~III] emission;
H and K refer to Ca II absorption;
G refers to G-band absorption; for the purposes of this table,
H$\alpha$, H$\beta$, H$\gamma$, and H$\delta$ refer to emission lines
H$\eta$ and H$\zeta$ refer to absorption lines.   }
\end{deluxetable}



\newpage

\begin{thebibliography}{}

\bibitem[Casertano, S., et al.\ (2000)]{cas00} Casertano, S., et
al. 2000, \aj, 120, 2747

\bibitem[Cohen, J.G., et al.\ (1996)]{coh96} Cohen, J.G., Cowie, L.L.,
Hogg, D.W., Songaila, A., Blandford, R., Hu, E.M., Shopbell,
P. \apjl, 475, L5

\bibitem[Ellison, Mall\'en-Ornelas \& Sawicki (2003)]{ems03} Ellison,
S.L., Mall\'en-Ornelas, G., \& Sawicki, M.\ 2003, ApJ, in press

\bibitem[Labb\'e, et al.\ (2003)]{lab03} Labb\'e, I., et al. 2003,
\aj, 125, 1107

\bibitem[Lowenthal et al.\ (1997)]{low97} Lowenthal, J.D., et
al. 1997, \apj, 481, 673

\bibitem[Oke 1974]{oke74} Oke, J.B. 1974, \apjs, 27, 21

\bibitem[Sawicki et al.\ (1997)]{saw97} Sawicki, M.J., Lin, H., \& Yee,
H.K.C. 1997, \aj, 113, 1

\bibitem[Steidel et al.\ (1996)]{ste96} Steidel, C.C., Giavalisco, M.,
Dickinson, M., \& Adelberger, K.L. 1996, \aj, 112, 352

\bibitem[Vanzella et al.\ (2002)]{van02} Vanzella, E., Cristiani, S.,
Arnouts, S., Dennefeld, M., Fontana, A., Grazian, A., Nonino, M.,
Petitjean, P., Saracco, P. 2002, \aap, 396, 847

\bibitem[Williams et al.\ 2000]{wil00} Williams, R.E., et al.\ 2000,
\aj, 120, 2735

\bibitem[Yee et al.\ (1991)]{yee91} Yee, H.K.C. 1991, \pasp, 103, 396

\end{thebibliography}
\end{document}